\def\papertitle{Deep Regularized RNNs for Virtual Analog Modeling}
\def\paperauthorA{V.~Valtteri Kallinen}
\def\paperauthorB{Lauri Juvela}
\def\paperauthorC{Thom Sherson}

\documentclass[twoside,a4paper]{article}
\usepackage{etoolbox}

\usepackage[print]{dafx26v2}

\usepackage{amsmath,amssymb,amsfonts,amsthm}
\usepackage{siunitx}
\usepackage{euscript}
\usepackage[T1]{fontenc}
\usepackage[utf8]{inputenc}
\usepackage{ifpdf}
\usepackage[english]{babel}
\usepackage{caption}
\usepackage{subfig} %
\usepackage{color}
\usepackage{booktabs}
\usepackage{lipsum}

\input glyphtounicode
\ninept

\newcounter{numauth}
\newcounter{listcnt}
\newcommand\authcnt[1]{\ifdefined#1 \stepcounter{numauth} \fi}

\newcommand\addauth[1]{
\ifdefined#1 
\stepcounter{listcnt}
\ifnum \value{listcnt}<\value{numauth}
\appto\authorslist{, #1}
\else
\appto\authorslist{~and~#1}
\fi
\fi}
\authcnt{\paperauthorB}
\authcnt{\paperauthorC}
\authcnt{\paperauthorD}
\authcnt{\paperauthorE}
\authcnt{\paperauthorF}
\authcnt{\paperauthorG}
\authcnt{\paperauthorH}
\authcnt{\paperauthorI}
\authcnt{\paperauthorJ}
\def\authorslist{\paperauthorA}
\addauth{\paperauthorB}
\addauth{\paperauthorC}
\addauth{\paperauthorD}
\addauth{\paperauthorE}
\addauth{\paperauthorF}
\addauth{\paperauthorG}
\addauth{\paperauthorH}
\addauth{\paperauthorI}
\addauth{\paperauthorJ}

\usepackage{times}

\newif\ifpdf
\ifx\pdfoutput\relax
\else
   \ifcase\pdfoutput
      \pdffalse
   \else
      \pdftrue
   \fi
\fi

\ifpdf %
  \usepackage[pdftex,
    pdftitle={\papertitle},
    pdfauthor={\authorslist},
    pdfsubject={Proceedings of the 29th International Conference on Digital Audio Effects (DAFx26)},
    colorlinks=false, %
    bookmarksnumbered, %
    pdfstartview=XYZ %
  ]{hyperref}
  \usepackage[pdftex]{graphicx}
\else %
  \usepackage[dvips]{epsfig,graphicx}
  \usepackage[dvips,
    pdftitle={\papertitle},
    pdfauthor={\authorslist},
    pdfsubject={Proceedings of the 29th International Conference on Digital Audio Effects (DAFx26)},
    colorlinks=false, %
    bookmarksnumbered, %
    pdfstartview=XYZ %
  ]{hyperref}
\fi
\usepackage[hypcap=true]{caption}
\title{\papertitle}

\twoaffiliations
{\paperauthorA\ and \paperauthorB}
{\href{https://www.aalto.fi/en/department-of-information-and-communications-engineering}{Dept. of Information and Communications Engineering (DICE)} \\ Aalto University \\ Espoo, Finland\\
{\tt \href{mailto:valtteri.kallinen@aalto.fi}{valtteri.kallinen@aalto.fi}}
}
{\paperauthorC}
{\href{https://neuraldsp.com}{Neural DSP Technologies Oy} \\ Helsinki, Finland\\
{\tt \href{mailto:thom@neuraldsp.com}{thom@neuraldsp.com}}
}

\usepackage{xurl}  %
\usepackage{microtype}
\usepackage[colorinlistoftodos, disable]{todonotes}
\usepackage{xspace}

\newcommand{\lstm}{$\text{LSTM}$\xspace}
\newcommand{\lstmtwo}{$\text{LSTM}_{2}$\xspace}
\newcommand{\lstminf}{$\text{LSTM}_{\infty}$\xspace}
\renewcommand{\vec}[1]{\mathbf{#1}}
\newcommand{\mat}[1]{\mathbf{#1}}
\newcolumntype{C}{@{\extracolsep{\fill}}c@{\extracolsep{0pt}}}

\graphicspath{{Pictures/}}

\begin{document}
\ifpdf %
  \DeclareGraphicsExtensions{.png,.jpg,.pdf}
\else  %
  \DeclareGraphicsExtensions{.eps}
\fi

\makeatletter
\newcommand{\ShowBaseSize}{%
  Base font size: \f@size pt, baseline skip: \the\baselineskip\par
}
\makeatother

\maketitle

\begin{abstract}

Virtual analog (VA) modeling methods seek to emulate analog audio hardware using digital signal processing (DSP). %
Modeling approaches fall into three broad categories: white-box methods, which use detailed device knowledge for accurate simulation; gray-box methods that use generic DSP blocks to model the system; and black-box methods, which rely solely on opaque models learned from input–output data. %
A category of architectures used widely in black-box modeling are recurrent neural networks (RNNs).  %
To model device controls, the control values can be provided as conditioning input to the network. %
However, when the conditioning is time-varied, the models are susceptible to producing noise artifacts. %
Regularization of the RNN dynamics significantly reduces these artifacts, though at a loss in modeling accuracy. %
This paper closes the dynamics regularization quality gap by introducing deep control-conditioned LSTMs and a gammatone filterband (GFB) loss. Experiments indicate that the proposed method achieves comparable modeling performance as unregularized baselines while avoiding the noise artifacts caused by time-varying control inputs. 
\end{abstract}

\section{Introduction}
\label{sec:intro}

Virtual analog (VA) modeling seeks to digitally emulate analog audio hardware,
enabling faithful reproduction of device behavior within modern signal-processing systems and workflows. 
Methodologically, VA modeling spans a spectrum from white-box
models that fully derive their behavior from circuit schematics through gray- and black-box models in which the desired behavior is inferred or learned from measured data. 
Gray-box models lie between the two extremes and typically employ traditional digital signal processing (DSP) blocks (filters, nonlinear waveshapers, dynamic processors) using some prior general model of the device behavior \cite{10.3389/frsip.2025.1580395}.
Black-box methods, on the other hand, typically leverage conventional machine learning (ML) architectures to capture the nonlinear and dynamic characteristics of audio circuits \cite{10.3389/frsip.2025.1580395}.
These data-driven models are especially useful when detailed expert knowledge or the time required for analyzing electronic circuits is unavailable.
In this work, we follow the black-box, ML-based line of research.

Recently, recurrent neural networks (RNNs), especially long short-term memory (LSTM) \cite{10.1162/neco.1997.9.8.1735} and gated recurrent unit (GRU) \cite{cho-etal-2014-properties} architectures, have successfully been applied to black-box virtual analog modeling applications \cite{10.3389/frsip.2025.1580395, app10020638, app10030766}.
Such networks are trained in a supervised manner using paired input and output audio recorded from a device.
Given enough data, the models learn to generalize to new inputs and produce an output that closely matches the target device.

When used to simulate devices that have controls to adjust various processing-related parameters (such as filter cutoffs, resonances, amount of distortion, or time constants), control values are commonly recorded as static parameter metadata with the audio data \cite{10094769, Mikkonen2024SamplingTU}.
These can be used as extra inputs to condition the model, allowing controllability over the sound as with the real hardware.
There are several proposed methods for applying control conditioning to RNNs: the simplest solution is to concatenate the control values with the audio input \cite{10094769, Mikkonen2024SamplingTU, Wright2019RealtimeBM}.
Another general approach is to directly modulate either the parameters of the model or the hidden features using an additional hyper-network \cite{simionato_2024_14361773, DAFx24_paper_16}.

In general, neural networks for black-box VA modeling are trained using a time or frequency-domain loss function, or a combination of the two \cite{8682805, Huhtala2024KLANNLL}.
The most frequently employed methods include the use of mean absolute error (MAE) \cite{DAFx24_paper_16}, mean squared error (MSE) \cite{simionato_2024_14361773}, error-to-signal ratio (ESR) \cite{8682805}, and multi-resolution short-time Fourier transform (MR-STFT) \cite{DAFx24_paper_16, Huhtala2024KLANNLL} or their mappings to other scales, such as the Mel scale \cite{DAFx25_paper_61}.

Time-domain loss functions are beneficial since they preserve the phase information of the target signal. 
Correct phase response is important in virtual analog modeling of distortion and compressor effects since these are commonly blended with the dry signal. 
A mismatch in phase response can result in unwanted amplification or cancellation in such cases. 
However, time-domain loss values map poorly to the perceptual quality of the output \cite{hu:2008:quality}, and therefore frequency-domain loss functions are advantageous when aiming for higher perceptual accuracy. 
Therefore, a common approach is to combine time and frequency-domain loss functions and adjust the sum to get the desired behavior. 
Unfortunately, this introduces another hyperparameter, and balancing multiple loss functions can be difficult since the time and frequency-domain losses often have differing scales and convergence rates.

Recently, in \cite{Kallinen:2025:AsymptoticallyStable}, it was demonstrated that LSTMs and GRUs suffer from audible artifacts caused by time-varying control conditioning.
To eliminate this behavior, the work proposed restricting the RNN dynamics to an asymptotically stable regime based on similar ideas in the fields of control and systems theory \cite{Stipanovi2018SomeLS, Deka2018GlobalAS, Bonassi2020OnTS, https://doi.org/10.1002/rnc.5519}.
However, the regularized models did not reach similar performance as their original unregularized counterparts.
Additionally, the study was restricted to only a single configuration of the LSTM and GRU networks, namely models with a single 32-unit RNN layer.

This paper extends the work in \cite{Kallinen:2025:AsymptoticallyStable} by introducing a deep control-conditioned architecture for VA modeling based on LSTMs. Furthermore, a gammatone filterbank (GFB) loss for model training is proposed. The results indicate that, when combined, the GFB loss and deep architecture enable the regularized models to match the performance of MAE-trained baselines while reducing conditioning-induced noise. It is further demonstrated that regularized models benefit from additional depth compared to simply increasing the layer width.

The remainder of the paper is organized as follows: Section~\ref{sec:methods} details the regularization formulations, the network architecture, and the GFB loss. Section~\ref{sec:experiments} describes the experimental setup, including tested configurations, datasets, and training procedures. Section~\ref{sec:results} presents the empirical results. Section~\ref{sec:discussion} provides discussion, and Section~\ref{sec:conclusions} concludes and outlines future directions.
Code used for training, sound demos of the models, and training results are provided online.
\footnote{\scriptsize \url{https://codeberg.org/rantlivelintkale/dr-rnn-va}}
\footnote{\scriptsize \url{https://rantlivelintkale.codeberg.page/dr-rnn-va/}}
\footnote{\scriptsize \url{https://doi.org/10.5281/zenodo.20406285}}

\section{Methods}
\label{sec:methods}

This section introduces the modeling approach, including the regularization applied to recurrent layers, the control-conditioning strategy for deep LSTMs, and the GFB augmentation for time‑domain loss.

\subsection{Regularized RNNs}

\begin{figure}[t]
    \centering
    \includegraphics[]{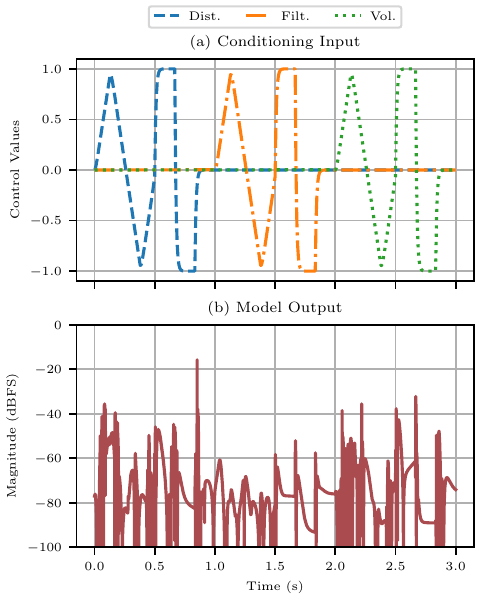}
    \caption{An example of typical control noise generated by a model under time-varying control conditioning with zero audio input. Output plotted for a $4 \times 64$ LSTM model trained on the RAT dataset. Varied control conditioning creates unwanted audible sound artifacts. Conditioning inputs are smoothed using a low-pass filter to produce values that resemble ideal conditions in a real-time use context.}
    \label{fig:decay}
\end{figure}

As demonstrated in \cite{Kallinen:2025:AsymptoticallyStable}, the key benefit of regularized RNNs is reduced control conditioning-induced noise that can happen during real-time inference when the control values are varied.
This is demonstrated in Figure\ \ref{fig:decay}: as the controls are varied, the model produces an audio output resembling a crackling sound. In \cite{Kallinen:2025:AsymptoticallyStable}, limiting the model's dynamic behavior under zero audio input to an asymptotically stable regime was found to eliminate this noise.

Let $\vec{x}_t$ denote the current audio input, $\vec{h}_t$ the hidden state, and $\vec{p}_t$ the conditioning input at time step $t$.
Additionally, let $\mat{W}_*$, $\mat{U}_*$, and $\mat{C}_*$ denote the input, hidden, and conditioning weight matrices, respectively, and let $\vec{b}_*$ denote the bias vectors.
Finally, notating the standard sigmoid logistic function using $\sigma(\cdot)$, the hyperbolic tangent function using $\phi(\cdot)$, and the Hadamard product using $\odot$; the operation of a control-conditioned LSTM layer is governed by the following set of equations:
\begin{align}
    \vec{i}_t &= \sigma(\mat{W}_i \vec{x}_t + \mat{U}_i \vec{h}_{t-1} + \mat{C}_i \vec{p}_t + \vec{b}_i), \\
    \vec{f}_t &= \sigma(\mat{W}_f \vec{x}_t + \mat{U}_f \vec{h}_{t-1} + \mat{C}_f \vec{p}_t + \vec{b}_f), \\
    \vec{g}_t &= \phi(\mat{W}_g \vec{x}_t + \mat{U}_g \vec{h}_{t-1} + \mat{C}_g \vec{p}_t + \vec{b}_g), \\
    \vec{o}_t &= \sigma(\mat{W}_o \vec{x}_t + \mat{U}_o \vec{h}_{t-1} + \mat{C}_o \vec{p}_t + \vec{b}_o), \\
    \vec{c}_t &= \vec{f}_t \odot \vec{c}_{t-1} + \vec{i}_t \odot \vec{g}_t, \\
    \vec{h}_t &= \vec{o}_t \odot \phi(\vec{c}_t).
\end{align}

This work employs the regularized LSTM variant from \cite{Kallinen:2025:AsymptoticallyStable}, where the following constraints were shown to guarantee asymptotic stability \cite{Kallinen:2025:AsymptoticallyStable}:
\begin{equation}
    \label{eq:lstm-inf}
    \mat{C}_g = O, \quad \vec{b}_g = \vec{0}, \quad \| \mat{U}_g \|_\infty < 1, \quad \| \vec{f}_{t} + \vec{i}_{t} \|_\infty \leq 1,
\end{equation}
where $\| \cdot \|_\infty$ denotes the induced matrix $L_\infty$-norm. Additionally, we conduct a comparison utilizing a less strict regularization for the hidden weights of the cell gate ($\mat{U}_{g}$) by replacing the  $L_\infty$-norm with the spectral norm (induced $L_2$-norm) such that
\begin{equation}
    \label{eq:lstm-2}
    \quad \| \mat{U}_g \|_2 < 1.
\end{equation}
While this does not formally satisfy the same stability guarantees as the $L_\infty$-norm variant, in practice it was found to still significantly reduce control noise while allowing slightly better use of model capacity.

Equations~(\ref{eq:lstm-inf}, \ref{eq:lstm-2}) were implemented via a reparametrization of the model parameters as in \cite{Salimans2016WeightNA,miyato2018spectral}.
Using reparameterizations means the model will always satisfy the constraints defined by the equations during and after training.
The reparametrization for the hidden weight matrix norms was implemented as follows:
\begin{equation}
    \mat{U}_{g} = \mat{U}_{g} \cdot
    \left\{
    \begin{array}{c l}
        \dfrac{\tau}{\| \mat{U}_{g} \|} & \text{if } \| \mat{U}_{g} \| \geq \tau , \\[12pt]
         1 & \text{otherwise} ,
    \end{array}
    \right.
\end{equation}
where $\tau < 1$ is a threshold to satisfy the constraint $\| \mat{U}_{g} \| < 1$.
The input gate $\vec{f}_t$ and forget gate $\vec{i}_{t}$ were reparameterized as in \cite{Kallinen:2025:AsymptoticallyStable} by mirroring forget gate parameters to the input gate, that is
\begin{equation}
    \mat{U}_{i} = -\mat{U}_{f}, \quad \mat{C}_{i} = -\mat{C}_{f}, \quad \vec{b}_{i} = -\vec{b}_{f},
\end{equation}
and limiting the parameter values of $\mat{U}_{f}$, $\mat{C}_{f}$ and $\vec{b}_{f}$ such that $\| \vec{f}_{t} \|_{\infty} \leq 1$.
Since mirroring these weights guarantees that $\vec{i}_{t} = \vec{1} - \vec{f}_{t}$ and thus $\| \vec{f}_{t} + \vec{i}_{t} \|_\infty \leq 1$ under zero audio input.

\subsection{Deep Architecture for Concatenation-Conditioned RNNs}

\begin{figure}[t]
    \centering
    \includegraphics[width=0.9\linewidth]{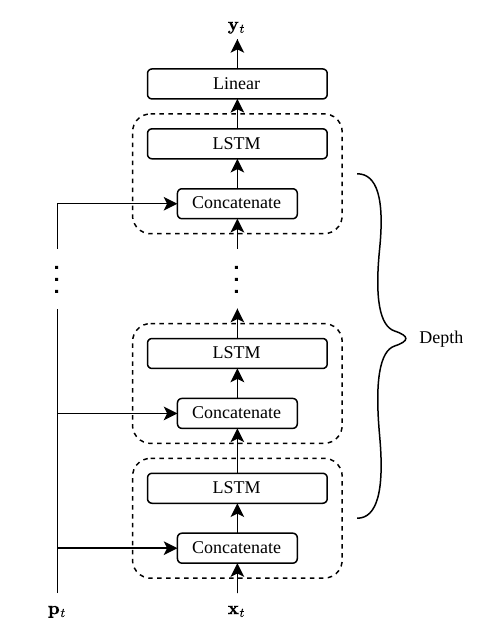}
    \caption{Deep architecture for control-conditioned LSTMs. The first layer receives as input the concatenation of audio signal $\vec{x}_{t}$ and control conditioning $\vec{p}_{t}$. Subsequent layers receive as input the concatenation of the previous layers hidden and the control conditioning $\vec{p}_{t}$. The output signal $\vec{y}_{t}$ is produced by passing the output of the final LSTM layer through a fully connected layer. Depth refers to the number of serial LSTM layers.}
    \label{fig:network}
\end{figure}

The deep conditioned LSTM network used in this paper is implemented as shown in Fig.~\ref{fig:network}. The network takes an audio signal $\vec{x}_{t}$ and conditioning input $\vec{p}_{t}$. The conditioning is applied by simply concatenating it as an extra input channel to each LSTM layer. This can be considered an extension of the common concatenation-based conditioning used in \cite{10094769, Mikkonen2024SamplingTU, Wright2019RealtimeBM} to deeper networks. The architecture is not LSTM specific; for example, the LSTM layers could be replaced by GRUs. Notably, regularized deep GRU networks can be implemented using the constraints for GRUs as defined in \cite{Kallinen:2025:AsymptoticallyStable}.

\subsection{Model Architecture and Regularization Notation}

The complete models with regularization are notated as follows:
\begin{equation}
    \label{eq:complete-notation}
    \mathrm{LSTM}_\mathrm{norm} \ \mathrm{depth} \times \mathrm{hidden\ size},
\end{equation}
where `norm' is either the spectral norm ($2$) Eq.~(\ref{eq:lstm-2}) or the infinity norm ($\infty$) Eq.~(\ref{eq:lstm-inf}), depth is the number of layers, and hidden size is the number of units in each LSTM layer.

\subsection{Gammatone Filterbank Extension for Time-Domain Loss}

In \cite{9052944}, the standard time-domain loss was combined with an A-weighting filter to achieve better perceptual quality without using a frequency-domain loss function. In this paper we adopt a similar perceptually inspired approach. First, the error signal between the target and prediction is calculated as usual:
\begin{equation}
    \vec{e}_{t} = \vec{y}_{t} - \hat{\vec{y}}_{t}.
\end{equation}
This error signal is then weighted using a gammatone filter bank (GFB) \cite{Patterson1988Gammatone}, the absolute errors of the channels are summed, and the mean of the summed errors at each time step is calculated.
The GFB-filtered error should provide better accuracy across multiple frequency ranges, similar to the common frequency-domain loss functions, while preserving the phase response since the loss is calculated from the time-domain error signal.

In this paper, a 32-channel filterbank is used with frequencies spread out over the ERB scale starting from a frequency of $\qty{100}{\hertz}$ with subsequent frequencies being a distance of one further on the ERB scale as defined in \cite{GLASBERG1990103}. With 32 channels, this gives us a GFB with frequencies up to $\approx \qty{10}{\kilo\hertz}$.
Since the ERB scale as defined in \cite{GLASBERG1990103} is only applicable for frequencies between $\qty{100}{\hertz}$ and $\qty{10}{\kilo\hertz}$, we include a residual error term, which is the difference between the filter outputs and the error signal, to model the error outside the GFB. Letting $C$ be the number of channels in the gammatone filterbank, this residual is defined as:
\begin{equation}\label{loss::gfb}
    \mathrm{GFB}(\vec{e}_{t})_{t,\text{res}} = \vec{e}_{t} - \sum_{c=1}^{C} \mathrm{GFB}(\vec{e}_{t})_{t,c}.
\end{equation}

Letting $N$ be the number of samples, the GFB loss is then defined as:
\begin{equation}
    \mathcal{L}_{\text{GFB}} = \frac{1}{N} \sum_{t=1}^{N} \left( | \mathrm{GFB}(\vec{e}_{t})_{t,\text{res}} | + \sum_{c=1}^{C} | \mathrm{GFB}(\vec{e}_{t})_{t,c} | \right).
\end{equation}

Other configurations with varying numbers of channels, with or without the residual, could also be applied. However, improved results were already observed using this 32-channel layout with the residual. Thus, this arrangement was used during this study.

\subsection{Evaluation Metrics}
\label{sec:metrics}

To compare the performance of the different architectures proposed, we consider three metrics: ESR \cite{app10030766, Wright2019RealtimeBM}, MR-STFT as defined and configured in \cite{9053795}, and the GFB loss described in the previous section.

In addition, we report the energy of the control conditioning noise generated by each model.
For these control noise (CN) tests, each network was first initialized using an impulse and allowed to stabilize for one second. Then, the controls were varied, and the signal energy during the time-varying conditioning was recorded. The controls used for conditioning were varied in the same way as in Fig.~\ref{fig:decay}.

\section{Experiments}
\label{sec:experiments}

This section details the experiments performed, including a description of the evaluated systems and data, the measurement and training procedures, and the model configurations used to study the effects of depth, width, and regularization.

\subsection{Testing Different Network and Loss Configurations}
\label{sec:tests}

Short training runs of 100 epochs each were conducted to investigate the way in which performance scaled with model width (number of units per RNN layer) and depth (number of sequential RNN layers). The tested configurations included all combinations of hidden sizes $d \in \{8, 16, 32, 64\}$ and depths $l \in \{1,2,3,4\}$. In addition, training was performed for both unregularized and regularized LSTM variants to detect any impacts that regularization may have on this scaling. These short runs were trained and evaluated using the GFB loss defined in Eq.~(\ref{loss::gfb}).

Based on the results of the short training runs, three configurations were chosen for longer training runs of 1000 epochs ($\approx \num{700000}$ parameter updates for all datasets) to allow the models to fully converge. The configurations used were one layer of $64$ units ($1 \times 64$), four layers of eight units ($4 \times 8$) and four layers of $64$ units ($4 \times 64$). Similarly, the unregularized and regularized variants were tested. Furthermore, to observe the effects of the GFB loss on performance, the models were also trained using the MAE loss as a baseline. These longer training runs were conducted using three different random number generator seeds to account for sampling bias.

\subsection{Datasets and Training Details}

\begin{figure*}[t]
    \centering
    \includegraphics[]{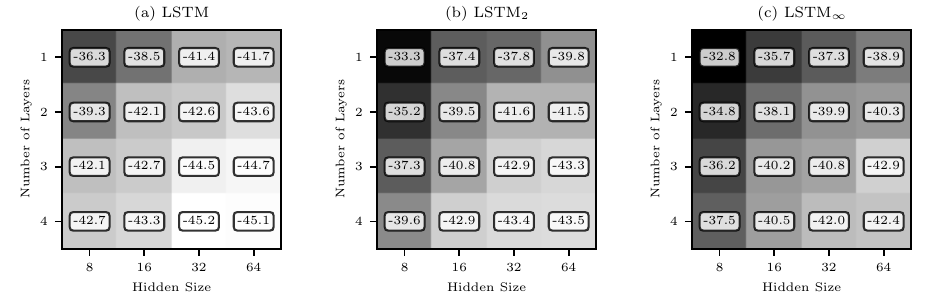}
    \caption{Training results from the short trial for different configurations of (a) LSTM, (b) \lstmtwo and (c) \lstminf on the RAT dataset. Each cell of a results matrix is a specific configuration of depth and width, for example, 4 layers of 8 LSTM units. Cell values indicate the best GFB evaluation loss on a decibel scale during a training run. Lower values are better.}
    \label{fig:result-matrix}
\end{figure*}

As in \cite{Kallinen:2025:AsymptoticallyStable}, this paper uses the ``Dataset for asymptotically stable recurrent neural networks'' \cite{kallinen_2026_20406285}.
The training dataset consists of three devices: the ProCo Rat (RAT), Darkglass Duality Fuzz (DFZ), and Boss CS-3.
The audio in the dataset consists of 1-second clips of dry electric guitar and bass guitar playing.
The dataset was captured using different control combinations, and therefore it can be used to train control-conditioned models. For the ProCo Rat, the dataset has values for all three controls: distortion, filter, and volume.
For the Duality Fuzz, the fuzz blend and filter controls were varied while the volume was set to the maximum position and the dry blend set to fully wet.
The CS-3 dataset only has values for the attack control, while the volume and sustain controls were set to their maximum values while the tone was set to the minimum value.
All audio uses a sampling rate of \qty{48}{\kilo\hertz} and the trained models operate under this sampling rate assumption for the input audio. The dataset is split into training and validation sets, and the validation set was used for testing as well.
Further details can be found in the dataset repository.

The reason for choosing to use this dataset is the current lack of high-quality, openly available datasets for control-conditioned training.
The Marshall JVM 410H dataset \cite{stepan:2023:marshall} contains too few control positions to allow a black-box model to properly generalize across different control position combinations.
The pOD dataset \cite{dalri2025morphdrive}, while very extensive, only includes recorded affected wet signals. Although the recorded signals were adjusted for AD/DA conversion latency, this does not account for any coloration or phase distortion caused by the passive re-amplification device used before the overdrive pedals. Moreover, all the pedals are variations of different overdrive pedals, and the controls are limited to two parameters: gain and tone.

The models and training were implemented using the PyTorch machine learning (ML) framework \cite{10.5555/3454287.3455008}.
The Adam optimizer was used for training with a learning rate of $10^{-3}$ and no weight decay.
As in \cite{Wright2019RealtimeBM}, the models were trained using truncated backpropagation through time (TBPTT) with a truncation step size of $2048$ samples and the training data was processed in batches of $32$ samples.

\section{Results}
\label{sec:results}

This section reports objective evaluation outcomes across model configurations, including results from the short runs and extended trainings, summary tables for evaluation metrics, parameter counts, and Bark‑smoothed error‑residual spectra over the evaluation sets.

\subsection{Results for Different Network Configurations}

Fig.~\ref{fig:result-matrix} shows the results for the network configuration test outlined in Sec.~\ref{sec:tests}. For all architectures, increasing model capacity by increasing depth and width resulted in a reduction in evaluation loss. However, in the case of the regularized variants, a degradation in model performance for equivalent network size can be noted. This impact is slightly reduced for the proposed $L_2$-norm regularization, although a gap with the standard LSTM persists.

For the longer training runs, also described in Sec.~\ref{sec:tests}, the results are gathered in Table~\ref{tab:final-losses}.
The table lists the metrics for the best models out of three runs.
Observing the results for the models shown in Fig.~\ref{fig:result-matrix} indicates that the models chosen in the first run still improved based on the evaluation GFB loss values.

\begin{table*}[t]
    \centering 
    \caption{Evaluation results for the best models from $3$ longer training runs (1000 epochs). Metrics were computed on the evaluation sets of each device, except for the control noise (CN) which was computed as described in \ref{sec:metrics}. Lower values indicate better performance. A decibel value of $-\infty$ indicates a value of exactly zero on a linear scale.}
    \resizebox{\linewidth}{!}{
    \begin{tabular*}{1.3\linewidth}{lll C cccc C cccc C cccc}
        \toprule
        &&&& \multicolumn{4}{c}{RAT} && \multicolumn{4}{c}{DFZ} && \multicolumn{4}{c}{CS-3} \\
        \cmidrule(){5-8} \cmidrule(){10-13} \cmidrule(){15-18}
        Loss &
        Model &
        Config. &&
        {$\text{ESR}_\text{dB}$} & {MR-STFT} & {$\text{GFB}_\text{dB}$} & {$\text{CN}_\text{dB}$} &&
        {$\text{ESR}_\text{dB}$} & {MR-STFT} & {$\text{GFB}_\text{dB}$} & {$\text{CN}_\text{dB}$} &&
        {$\text{ESR}_\text{dB}$} & {MR-STFT} & {$\text{GFB}_\text{dB}$} & {$\text{CN}_\text{dB}$} \\
\midrule
MAE&\lstm&$1 \times 64$&&${-17.2}$&${0.248}$&${-43.6}$&${-64.2}$&&${-19.0}$&${0.449}$&${-31.1}$&${-37.4}$&&$\mathbf{-27.1}$&$\mathbf{0.080}$&${-63.7}$&${-74.4}$\\
&&$4 \times 8$&&${-15.2}$&${0.288}$&${-42.4}$&${-65.2}$&&${-15.2}$&${0.563}$&${-27.3}$&${-26.9}$&&${-23.4}$&${0.111}$&${-59.9}$&${-78.7}$\\
&&$4 \times 64$&&${-18.9}$&${0.198}$&${-45.1}$&${-65.0}$&&${-20.3}$&${0.442}$&${-33.3}$&${-33.5}$&&${-21.4}$&${0.133}$&${-63.9}$&${-58.0}$\\
\cmidrule{2-18}
&\lstmtwo&$1 \times 64$&&${-16.4}$&${0.343}$&${-42.0}$&${-\infty}$&&${-15.1}$&${0.548}$&${-26.1}$&${-\infty}$&&${-20.6}$&${0.133}$&${-56.6}$&${-209.1}$\\
&&$4 \times 8$&&${-16.0}$&${0.267}$&${-41.8}$&${-\infty}$&&${-17.0}$&${0.463}$&${-28.9}$&${-\infty}$&&${-21.3}$&${0.118}$&${-57.8}$&${-220.1}$\\
&&$4 \times 64$&&${-18.5}$&${0.211}$&${-44.8}$&${-\infty}$&&${-20.4}$&${0.404}$&${-32.4}$&${-\infty}$&&${-25.3}$&${0.090}$&${-63.6}$&${-156.8}$\\
\cmidrule{2-18}
&\lstminf&$1 \times 64$&&${-14.2}$&${0.403}$&${-39.6}$&${-\infty}$&&${-12.7}$&${0.641}$&${-23.3}$&${-\infty}$&&${-13.7}$&${0.247}$&${-50.6}$&${-\infty}$\\
&&$4 \times 8$&&${-15.4}$&${0.295}$&${-41.2}$&${-\infty}$&&${-17.0}$&${0.477}$&${-28.4}$&${-\infty}$&&${-19.1}$&${0.147}$&${-56.2}$&${-448.5}$\\
&&$4 \times 64$&&${-17.5}$&${0.226}$&${-44.0}$&${-\infty}$&&${-19.8}$&${0.423}$&${-32.1}$&${-\infty}$&&${-23.9}$&${0.102}$&${-62.5}$&${-\infty}$\\
\midrule
GFB&\lstm&$1 \times 64$&&${-17.6}$&${0.244}$&${-45.2}$&${-61.2}$&&${-18.9}$&$\mathbf{0.392}$&${-32.3}$&${-40.5}$&&${-25.9}$&${0.086}$&$\mathbf{-64.7}$&${-86.8}$\\
&&$4 \times 8$&&${-13.8}$&${0.341}$&${-43.2}$&${-57.1}$&&${-15.5}$&${0.493}$&${-29.8}$&${-28.8}$&&${-22.2}$&${0.115}$&${-60.2}$&${-64.1}$\\
&&$4 \times 64$&&$\mathbf{-19.3}$&$\mathbf{0.194}$&$\mathbf{-46.2}$&${-56.1}$&&${-19.0}$&${0.439}$&${-33.2}$&${-38.2}$&&${-25.6}$&${0.098}$&${-63.0}$&${-58.6}$\\
\cmidrule{2-18}
&\lstmtwo&$1 \times 64$&&${-17.3}$&${0.279}$&${-43.0}$&${-\infty}$&&${-15.8}$&${0.505}$&${-27.6}$&${-\infty}$&&${-20.1}$&${0.132}$&${-57.0}$&${-174.2}$\\
&&$4 \times 8$&&${-17.4}$&${0.255}$&${-43.4}$&${-\infty}$&&${-19.1}$&${0.431}$&${-31.1}$&${-\infty}$&&${-20.8}$&${0.126}$&${-57.7}$&${-165.4}$\\
&&$4 \times 64$&&${-19.2}$&${0.198}$&${-45.6}$&${-\infty}$&&$\mathbf{-20.8}$&${0.393}$&$\mathbf{-33.7}$&${-\infty}$&&${-26.5}$&${0.083}$&${-64.3}$&${-202.9}$\\
\cmidrule{2-18}
&\lstminf&$1 \times 64$&&${-15.7}$&${0.337}$&${-41.3}$&${-\infty}$&&${-12.8}$&${0.591}$&${-24.4}$&${-\infty}$&&${-13.3}$&${0.234}$&${-50.9}$&${-\infty}$\\
&&$4 \times 8$&&${-16.0}$&${0.283}$&${-41.9}$&${-\infty}$&&${-17.4}$&${0.465}$&${-29.7}$&${-\infty}$&&${-19.4}$&${0.142}$&${-56.6}$&${-\infty}$\\
&&$4 \times 64$&&${-18.0}$&${0.219}$&${-44.9}$&${-\infty}$&&${-19.9}$&${0.425}$&${-32.5}$&${-\infty}$&&${-23.2}$&${0.104}$&${-62.6}$&${-\infty}$\\

    \bottomrule
    \end{tabular*}
    } %
    \label{tab:final-losses}
\end{table*}

\subsection{Effect of Loss Function on Model Performance}

A general improvement in the evaluation ESR, MR-STFT, and GFB can be seen for models trained using the GFB loss. However, sometimes an improvement in GFB does not result in an improvement in ESR or MR-STFT, as is the case for some unregularized \lstm $4 \times 8$ models. For example, the MAE-trained \lstm $4 \times 8$ RAT model has an evaluation $\text{ESR}_\text{dB}$ of $-15.2$ while the same metric for the GFB-trained model is $-13.8$.

\subsection{Error Signal Spectrum Comparison}

\begin{figure*}[t]
    \centering
    \includegraphics[]{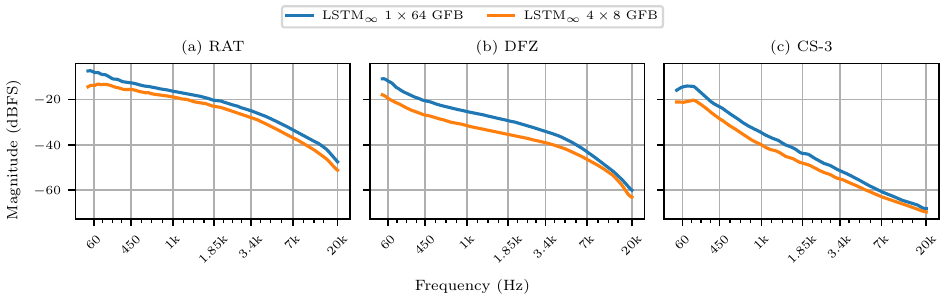}
    \caption{Bark smoothed spectrums of the error residuals over the evaluation sets for \lstminf $1 \times 64$ and $4 \times 8$ trained using GFB loss. Lower magnitudes indicate better performance. Deeper and narrower configurations consistently show better performance while having fewer parameters}
    \label{fig:spec1}
\end{figure*}

\begin{figure*}[t]
    \centering
    \includegraphics[]{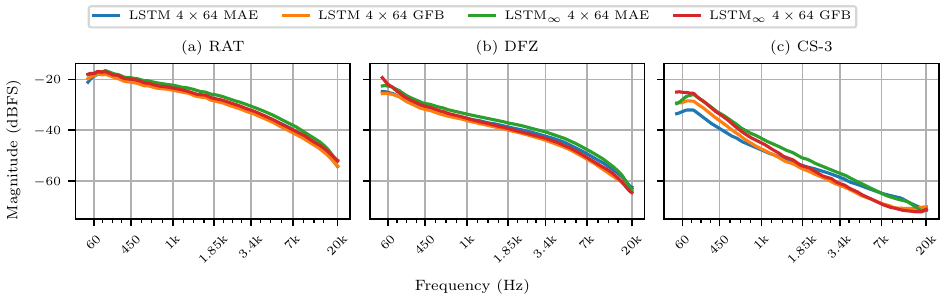}
    \caption{Bark smoothed spectrums of the error residuals over the evaluation sets for \lstm trained using the MAE loss and \lstminf models trained using MAE and GFB loss. Lower magnitudes indicate better performance. GFB loss helps the regularized model obtain similar performance to the unregularized baseline.}
    \label{fig:spec2}
\end{figure*}

The difference in modeling performance across the frequency spectrum is visualized in Fig.~\ref{fig:spec1} and \ref{fig:spec2}.
The figures depict the Bark-smoothed spectrums of the error residual from the evaluation set for each device.

Fig.~\ref{fig:spec1} shows the error residual spectrums for the $L_{\infty}$ regularized models for the $1 \times 64$ and $4 \times 8$ configurations.
It can be noted that the deeper and narrower $4 \times 8$ model consistently achieves lower error across the entire spectrum for all datasets.
The decreasing slope towards higher frequencies can be considered a reflection of the modeled device's output spectra and does not necessarily indicate a better modeling performance at higher frequencies.
However, the figures still highlight a relative improvement between the error spectrums of the models considered.
For the RAT, the results are improved evenly across the spectrum, whereas for the DFZ and the CS-3 the improvements drop off towards the higher frequencies.
A similar comparison was used in \cite{9052944} where an improvement in the error residual around sensitive frequencies correlated with better results during a listening test.
Thus, improvements in this metric should also indicate improvements in subjective results.

Fig.~\ref{fig:spec2} shows the error residuals for the MAE-trained \lstm and \lstminf and the GFB-trained \lstminf $4 \times 64$ models.
This shows the difference between the GFB loss-trained regularized model and the unregularized model, which can be considered a baseline for comparison.
While the difference is modest, the GFB loss enables the $L_\infty$-regularized model to reach similar performance as the unregularized counterpart. 

\subsection{Control Noise}

The results of the control noise test can be seen in Table~\ref{tab:final-losses}. The results indicate that models trained using the $L_2$-norm exhibit similar control noise suppression as the $L_\infty$-norm regularized models.
Since levels below $-100 \text{ dBFS}$ are lower than the noise floor for a generic audio interface, the control noise induced for the regularized models is practically zero under zero audio input.
That is, the highest control noise signal energy for the regularized models is $-156 \text{ dBFS}$.

\subsection{Computational Complexity}

To give a rough estimate of the computational complexity of the different configurations, Table~\ref{tab:param-counts} lists the number of parameters in the models trained on each dataset for the different configurations used in the longer training. The number of parameters scales quadratically with width while only scaling linearly with depth. As such, the $4 \times 8$ has the least number of parameters while the $4 \times 64$ has the most. Notably, the $4 \times 8$ models have approximately an order of magnitude fewer parameters than their $1 \times 64$ counterparts.

\begin{table}[t]
    \centering
    \caption{Model parameter amounts for different models and datasets. The RAT had three controls, the DFZ two controls, and the CS-3 one control. The number of parameters affects the size of the conditioning input vector and therefore the size of the conditioning weight matrices. The number of parameters does not significantly change with the number of conditioning inputs. The number of output parameters is the same for all datasets and listed separately. The floating-point operations (FLOPS) per sample approximation does not include nonlinear activations.}
    \resizebox{\linewidth}{!}{
    \begin{tabular}{l l l l l l}
        \toprule
        & \multicolumn{4}{c}{Number of Parameters} \\
        \cmidrule(lr){2-5}
        Config. & RAT & DFZ & CS-3 & Output Layer & FLOPS/sample \\
        \midrule
        $1 \times 64$ & \num{17 664} & \num{17 408} & \num{17 152} & \num{65} & $\approx \num{18 000}$ \\
        $4 \times 8$  & \num{2336} & \num{2208} & \num{2080} & \num{9} & $\approx \num{2400}$ \\
        $4 \times 64$ & \num{119 040} & \num{118 016} & \num{116 992} & \num{65} & $\approx \num{120 000}$ \\
        \bottomrule
    \end{tabular}
    }
    \label{tab:param-counts}
\end{table}

\section{Discussion}
\label{sec:discussion}

The results demonstrate a consistent trade-off between accuracy, capacity, and stability. Scaling depth and width lowers evaluation loss across architectures, confirming that larger models fit the data better; however, for equivalent sizes, both stability-regularized variants show degraded performance relative to the unregularized LSTM, with a smaller gap for the proposed $L_2$-norm regularization than for the $L_\infty$-constrained model. These patterns were first observed in the short, 100-epoch runs over depth and width, and motivated the choice of three representative configurations, $1 \times 64$, $4 \times 8$, and $4 \times 64$, for extended, 1000-epoch training runs.
With extended training, the selected models continued to improve, and additionally, the performance gap between unregularized and regularized models was narrowed. For example, the GFB loss between the $4 \times 8$ RAT models for the initial run was $\approx \qty{5.2}{\decibel}$ while after the longer training run, the difference was reduced to $\approx \qty{1.3}{\decibel}$.

Choice of training objective (MAE vs. GFB) affected the results in a generally consistent manner.
Comparing models trained with MAE versus the GFB loss, evaluation ESR, MR-STFT, and GFB metrics generally improved with the GFB objective, although gains in GFB did not always coincide with improvements in ESR and MR-STFT (notably for some $4 \times 8$ unregularized LSTMs).
Lower GFB loss values indicated better error spectra, which, given prior work in perceptual time-domain loss functions \cite{9052944}, should indicate subjective improvements in sound quality.

Architectural allocation of capacity also mattered under stability constraints. The results indicate that, for regularized models, deeper, narrower configurations ($4 \times 8$) outperform shallower, wider ones ($1 \times 64$) across devices. These findings, when combined with the way the number of parameters scales with depth versus width, suggest that regularized models benefit more from distributing capacity across depth.
For example, for the \lstminf $1 \times 64$ and $4 \times 8$ models trained on the DFZ dataset, the $\text{ESR}_\text{dB}$ values were $-12.8$ and $-17.4$, respectively. Whereas for the unregularized \lstm, the values for the same configurations were $-18.9$ and $-15.5$, indicating better utilization of the added depth despite the decrease in the total number of parameters.

\section{Conclusions}
\label{sec:conclusions}

This work shows that deeper, concatenation-conditioned LSTMs trained with the proposed GFB loss can achieve competitive modeling performance while eliminating the control-induced crackling artifacts that afflict unregularized models during real-time parameter changes. Across configurations, increasing capacity improves accuracy, but stability constraints introduce a modest performance penalty at matched sizes; the proposed $L_2$-norm regularization narrows this gap relative to the proven $L_\infty$-norm while empirically preserving the substantial control noise suppression.

Future work could pursue analytical stability characterizations for spectral-normalized recurrent layers and broaden empirical validation across additional devices and conditioning schemes.
Alternatively, with the knowledge of this study, one possible avenue of research is developing alternative architectures that inherently benefit from lower levels of control conditioning-induced noise.
Finally, training on time‑varying control trajectories rather than only static positions could be investigated to see if these effects are simply caused by overfitting.
However, currently there are no datasets with time-varying controls, and collecting such datasets presents many challenges related to the inherent difficulty in accurately actuating the controls.

\section{Acknowledgments}
We acknowledge the computational resources provided by the Aalto Science-IT project.
This research was funded by the Research Council of Finland (grant no. 371845).

\bibliographystyle{IEEEtranDAFx}
\bibliography{DAFx26_tmpl} %

\end{document}